\documentclass[twoside]{article}
\usepackage{qic,epsfig,amsmath,hyperref}
\usepackage[latin1]{inputenc}
\usepackage[OT1]{fontenc}
\newcommand{\beq}{\begin{equation}}
\newcommand{\bea}{\begin{eqnarray}}
\newcommand{\eea}{\end{eqnarray}}
\newcommand{\ba}{\begin{array}}
\newcommand{\ea}{\end{array}}

\newcommand{\bef}{\begin{figure}}
\newcommand{\eef}{\end{figure}}

\newcommand{\ie}{{\it i.e.\/}\ }
\newcommand{\eg}{{\it e.g.\/}\ }

\newcommand{\variance}[1]{\langle{#1}^2\rangle}
\newcommand{\condvar}[2]{V_{#1 \vert #2}}

\begin{document}
\setlength{\textheight}{8.0truein}    

\runninghead{Virtual Entanglement and Reconciliation Protocols for Quantum
            Cryptography with Continuous Variables}
            {F. Grosshans, N. J. Cerf, J.  Wenger, R. Tualle-Brouri 
              and Ph. Grangier}
\normalsize\textlineskip
\thispagestyle{empty}
\setcounter{page}{1}

\copyrightheading{0}{0}{2003}{000--000}

\vspace*{0.88truein}

\alphfootnote

\fpage{1}

\centerline{\bf
VIRTUAL ENTANGLEMENT AND RECONCILIATION PROTOCOLS}
\vspace*{0.035truein}
\centerline{\bf FOR QUANTUM CRYPTOGRAPHY WITH CONTINUOUS VARIABLES}
\vspace*{0.37truein}
\centerline{\footnotesize
FR\'ED\'ERIC GROSSHANS and NICOLAS J. CERF}
\vspace*{0.015truein}
\centerline{\footnotesize\it \'Ecole Polytechnique, CP 165, Universit\'e Libre de Bruxelles}
\baselineskip=10pt
\centerline{\footnotesize\it B-1050 Brussels, Belgium}
\vspace*{10pt}
\centerline{\footnotesize 
J\'ER\^OME WENGER, ROSA TUALLE-BROURI and PHILIPPE GRANGIER}
\vspace*{0.015truein}
\centerline{\footnotesize\it Laboratoire Charles Fabry de l'Institut d'Optique}
\baselineskip=10pt
\centerline{\footnotesize\it F-91403 Orsay cedex, France}
\vspace*{0.225truein}
\publisher{(received date)}{(revised date)}

\vspace*{0.21truein}

\abstracts{
  We discuss quantum key distribution protocols using
  quantum continuous variables. We show that such protocols can be made
  secure against individual gaussian attacks 
  regardless the transmission of the optical line between Alice and Bob.
  This is achieved by reversing the reconciliation procedure subsequent to the 
quantum transmission, that is, using Bob's instead of Alice's data to build the key. 
Although squeezing or entanglement may be helpful
to improve the resistance
  to noise, they are not required for the protocols to remain secure with high losses. 
Therefore, these protocols can be implemented very simply by transmitting coherent states and
  performing homodyne detection. }{ Here, we show that entanglement nevertheless
plays a crucial role in the security analysis of coherent state protocols. Every cryptographic 
protocol based on displaced gaussian states turns out to be equivalent to an 
entanglement-based protocol, even though no entanglement
is actually present.  This equivalence even holds in the absence of squeezing, 
for coherent state protocols. This ``virtual''
entanglement is important to assess the security of these protocols as it 
provides an upper bound on the mutual information between
Alice and Bob if they \emph{had} used entanglement. The resulting security 
criteria are compared to the separability criterion for
bipartite gaussian variables. It appears that the security thresholds are 
well within the entanglement region.
This supports the idea that coherent state quantum cryptography may
be unconditionally secure.}{}

\vspace*{10pt}

\vspace*{3pt}
\communicate{to be filled by the Editorial}

\vspace*{1pt}\textlineskip    
\section{Introduction}
\label{intro}

\subsection{Continuous-variable quantum cryptography}
In the presently very active field of continuous variable quantum information
processing, a stimulating question is whether quantum continuous variables
(QCV) \cite{braunstein-pati} may provide a valid alternative to the usual ``single photon" 
quantum key distribution (QKD) schemes \cite{gisin-rmp}.  
Many recent proposals to use QCV for QKD (for
a short review see \cite{prl}) have been based upon the use of ``non-classical"
states, such as squeezed or entangled light beams. We have nevertheless 
shown \cite{prl}, and experimentally demonstrated \cite{GVAWBCG03}, that
there is actually no need for squeezed  or entangled light: QKD can be implemented simply 
by generating and transmitting random distributions of
coherent states. More precisely, coherent state protocols are secure
against individual gaussian attacks, while their security with respect
to the line transmission depends on the {\it reconciliation protocol}
which is used by Alice and Bob to correct the transmission errors.
Using the so-called ``direct reconciliation'' (DR) protocols, a whole
family of secure protocols can be obtained by using either coherent states,
squeezed states, or Einstein-Podolsky-Rosen \cite{epr} (EPR) entangled beams
\cite{prl,CLVA00,CIVA01}, provided that the transmission of the line is 
larger than 50 percent (\ie the losses 
are less than 3 dB).  The security of these protocols is related to the limit 
imposed on the cloning of gaussian states \cite{cerf,cerf2,GG01}, so that 
non-classical features like squeezing or EPR correlations
have no influence on the achievable secret key rate.  Interestingly, the 
3~dB loss limit of these cryptographic protocols may be
circumvented by modifying the reconciliation  protocol. In ref. \cite{GVAWBCG03,GG02}, 
we have introduced   ``reverse reconciliation"
(RR) protocols, and demonstrated their security  for any value of the line transmission. 
Note that there exist, in principle, other ways for Alice and Bob
to go beyond the 3 dB limit of DR protocols, namely
by using entanglement purification \cite{Dal00}
or postselection \cite{silb}.

In the present paper, we will first review some basic properties of 
the direct and reverse reconciliation protocols. 
Then, we will show that each prepare-and-measure continuous-variable protocol is 
equivalent to an entanglement-based QKD protocol. This equivalence reminds us 
the link between the entanglement-free BB84 protocol
\cite{BB84} and the EPR-based protocol proposed by Ekert \cite{Eke91} that was 
pointed out in \cite{BBM92}.  This equivalence allows us
to compute the best estimate Alice may have on Bob's measurement outcome, 
if she \emph{had} used an entanglement-based protocol. This,
in turn, allows us to upper bound the information that an eavesdropper, Eve, 
can have on Bob's measurement results. In the case of a
channel with losses but no added noise, Eve's estimate turns out to be always 
worse than Alice's estimate, which is the main reason for
the increased security achieved by reversing the reconciliation protocol. 
Finally we will compare the security  criteria derived from
our approach to the entanglement criterion for bipartite gaussian variables. 
It appears that the corresponding security thresholds are
well within the entanglement region, supporting the idea that coherent states 
quantum cryptography may be unconditionally secure.

\subsection {Direct and reverse reconciliation protocols}

In the first step of a generic QKD protocol, Alice prepares a quantum state
and sends it to Bob, who makes a measurement on the state.  Alternatively,
Alice and Bob may share two EPR-correlated systems and both make a
measurement on their part.  In order to warrant security, Alice and Bob must
randomly choose to use different measurement bases, the transmitted data
being kept only when the bases are compatible.  After the quantum
exchange, they thus have to agree on a common measurement basis, and discard
the wrong measurements.  At the end of this step, Alice, Bob, and the
potential eavesdropper Eve, share a set of correlated data, called ``key elements''.

In a second step, Alice reveals some randomly chosen sample of the data that
she sent, and Bob reveals his corresponding measurements.  These samples allow
them to measure some relevant parameters of the quantum channel, \eg the error
rate and the transmission (called ``channel gain'' for QCV protocols).
Knowing the correlations between their key elements, Alice and Bob can
evaluate the amount of information they share ($I_{AB}$), and the information
the eavesdropper Eve may have at most about their key elements ($I_{AE}$ and $I_{BE}$). 
Therefore they can evaluate the size of the secret key they will be able 
to generate at the end of the protocol.  If Eve knows too
much, the size of this secret key will be zero, and Alice and Bob abort the protocol at this point.

In a third step, called ``reconciliation'', Alice and Bob use
classical communications to extract a common binary key from their correlated key elements, 
revealing as little information as possible to a third party
ignoring these key elements. This step usually uses parity-based algorithms
like Cascade. It was adapted to continuous variables in Refs.~\cite{CIVA01,VACC01}, 
where a ``sliced'' error correction procedure was devised in order
to provide reconciled bits from real-values key elements.
There are actually two main options for doing the
reconciliation, depending on whether Alice's or Bob's data are used to build the key. 
We will call these two options ``direct reconciliation'' (DR)
and ``reverse reconciliation'' (RR), respectively, and will detail these procedures 
in Sections \ref{sectDR} and \ref{sectRR}. The starting point will
be the Csiszar-K\"orner theorem \cite{CK78,Maurer} stating that a sufficient condition 
for distilling a secret key is that $\max(I_{AB}-I_{AE},I_{AB}-I_{BE})>0$,
the first and second term corresponding to DR and RR, resepctively.

Finally, the fourth step of a practical QKD protocol consists in
Alice and Bob performing ``privacy amplification'' in order to filter out Eve's information. 
Since this step is based on an evaluation of the amount of information collected by Eve on the
reconciled key, a crucial requirement is to get a bound on $I_{AE}$ for DR,
or on $I_{BE}$ for RR. For a coherent state protocol, the DR bound was given
in ref. \cite{prl}, and leads to a security limit for line with a transmission of 1/2. 
In the following, we will establish the RR bound and show that it is not associated 
with a minimum value of the line transmission.
In order to have a general approach, we will start by considering the exchange
of entangled beams, and we will show later that for a particular choice of 
the measurement performed by Alice, this is equivalent to exchanging coherent states.

\subsubsection{Direct Reconciliation (DR).}
\label{sectDR} 
In direct reconciliation, Alice sends correction information to Bob, who accordingly 
corrects his key elements to have the same values as Alice.  Alice infers from her estimate of $I_{AB}$ the minimum amount of
information she needs to reveal at this step.  If the reconciliation protocol is perfect, it keeps $I_{AB}-I_{AE}$ constant.  After
reconciliation, Alice and Bob know a common bit string of length $I_{AB}$ (slightly less if the reconciliation protocol is not
perfect), and Eve knows $I_{AE}$ bits of this string.  It will provide a usable secret key if $I_{AB}-I_{AE} > 0$.  We call this
``direct reconciliation'' (DR) because Bob is reconstructing what was sent by Alice, and the classical information flow in this step
has the same direction as the initial quantum information flow.
    
Direct reconciliation is quite intuitive, and it was used in the coherent
state QCV protocol that we proposed in ref. \cite{prl}. However, it is not
secure as soon as the quantum channel transmission falls below 1/2.
Intuitively, Eve could simulate the losses by a beam splitter and look one
output port of this beamsplitter. It seems obvious that, if she keeps the
biggest part of the beam sent by Alice (\ie if she simulate losses higher than
3~dB), she can extract more information from her beam than Bob
($I_{AE}>I_{AB}$), thus forbidding any secret key generation.  

Note that this limitation is actually not specific to QCV: a ``direct'' version of BB84 would be a protocol 
where Bob would try to fill in the ``empty slots'' where he did not get any photon. Such a  protocol actually only works when the losses
are smaller than 3~dB. Indeed, suppose Alice has a perfect photon-gun and sends single photons to Bob, who measures their polarization with
perfect detectors.  If $G<1$ denotes the transmission of the errorless lossy channel,
Bob only receives and measures a fraction $G$ of these photons. Even if we
suppose that Bob has a quantum memory, allowing him to always make the right basis choice, 
we have $I_{AB}=G$. If the losses are due to Eve, which keeps the lost photons, $I_{AE}=1-G$. The security condition $I_{AB}-I_{AE}>0$ for a
``direct'' version of BB84 is therefore $G>\frac12$. The usual BB84 protocol works for higher losses because only the photons received by Bob
(and therefore not intercepted by Eve) are considered for the key. As we will show in Sect. \ref{sectRR}, this may be viewed as a reverse
reconciliation where Alice corrects her value to match the ternary digit (0,1,no photon) held by Bob.

\subsubsection{Reverse Reconciliation (RR).}
\label{sectRR}

We may instead reverse the reconciliation in the sense that Bob
sends the correction information while Alice corrects her key elements to have the same values as Bob.  
Since Bob gives the correction information (also to Eve), this type of reconciliation keeps $I_{AB}-I_{BE}$ constant, and provides a
usable key if $I_{AB}-I_{BE} > 0$.  We call it ``reverse reconciliation''
(RR) because Alice adapts herself to what was received by Bob.

In a noiseless BB84 with finite line transmission, this step corresponds to
Bob informing Alice of his ``empty slots'' where he did not get any photon, and
Alice discarding the corresponding bits in order to have the same key.  In our
QCV protocol, there is no ``empty slot'' since homodyning the vacuum gives a gaussian distribution, and the RR procedure is intertwined
with error correction. Then, alike BB84, it allows Alice and Bob to cross
the 3-dB loss limit and extract a secret key for an arbitrarily low value 
of the line transmission.

However, in a practical realization, one cannot attain very high losses
for several reasons. First, a realistic reconciliation protocol cannot
reach the Shannon limit, so Alice and Bob actually obtain only a fraction
of the information $I_{AB}$ while one has to assume that Eve gets the full
information $I_{BE}$. Said otherwise, the correction information that must
be sent by Bob to Alice (but which is also monitored by Eve) is slightly larger than its ideal value 
predicted by Shannon theory. This makes the information
difference vanish at some finite value of the line transmission. Another
problem which must be taken into account is the following: while the RR procedure should be unidirectional 
(from Bob to Alice), the error correction using Cascade is a bidirectional process, so that some information also ``leaks'' from Alice to Eve. 
We have numerically evaluated this information leakage in practical cases \cite{GVAWBCG03} and it appears to be small, so we will not consider it
further in the present paper. However, it must be kept in mind that the one-way or two-way character of the used error correction procedure plays
a role, which should not be underestimated.

\section{Preparation of a modulated gaussian beam through entanglement}

The QKD protocols of the references \cite{prl,GVAWBCG03,CLVA00,CIVA01,GG02} are based on 
randomly displaced squeezed or coherent states prepared by
Alice. We will show in this section that Alice could equivalently prepare a
pair of quantum entangled beams, measure one (or both) quadratures on one beam, and send the other beam to Bob. 
This will be used in Sections~\ref{ReverseCloning} and \ref{security} to find the maximum information 
Alice may have on Bob's data if she was using
quantum entangled beams, and in Sect.~\ref{entanglement-security}
to compare the security conditions with the entanglement criterion for bipartite gaussian states.

\subsection{Measurement of a single quadrature}

\begin{figure} [tbp]
\centerline{\epsfig{file=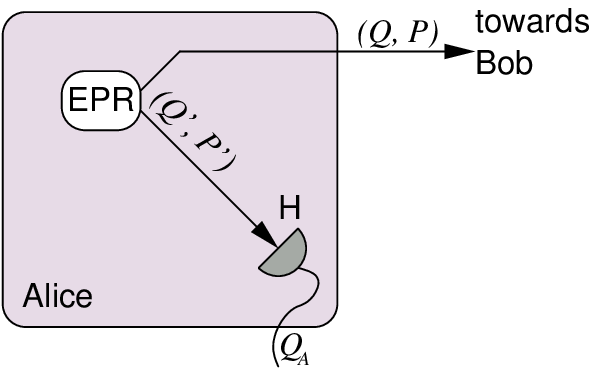, width=8.2cm}} 
\vspace*{13pt} \fcaption{\label{EPR_RR1}\textbf{Measurement of a single
    quadrature.}  \textsf{Alice} prepares two entangled beams $(Q,P)$ and
  $(Q',P')$ using an EPR source (\textsf{EPR}). She measures one
  quadrature on one beam with an homodyne detector (\textsf{H}) and deduces from it $Q_A$ (or $P_A$), 
which is an estimate of $Q$ (or $P$). She sends the other beam to \textsf{Bob}.}
\end{figure}

Let us assume Alice prepares a pair of EPR beams, and denote by $(Q, P)$ the
quadratures of the beam sent to Bob and by $(Q', P')$ the quadratures of the
beam kept by Alice (see Fig.~\ref{EPR_RR1}). To simplify the notations, we
will suppose those beams to be initially symmetric in the two quadratures, \ie
\begin{align}
\label{eq:VarEPR}
\variance{Q}=\variance{Q'}&=
V\,N_0 &
\variance{P}=\variance{P'}&= V\,N_0,
\end{align}
where $N_0$ is the shot-noise variance. 

These beams are entangled, and the measurement of a quadrature of one beam
(\eg $Q'$) gives Alice information on the same quadrature of the other
beam ($Q$).  One can show \cite{qnd,qnd2} that the best estimate Alice can have on $Q$ knowing $Q'$ 
is of the form $Q_A=\alpha Q'$ with $\alpha=\frac{\langle
  Q\,Q'\rangle}{\variance{Q'}}$, the value of $\alpha$ being found by
minimizing the variance of the error operator $\delta Q_A=Q-Q_A$.  The \emph{conditional variance}
$\condvar{Q}{Q_A}$ of $Q$ knowing $Q_A$ quantifies the remaining uncertainty on
$Q$ after the measurement of $Q'$ giving the estimate $Q_A$ of $Q$, and we have
\begin{equation}
  \label{eq:ValCondVar}
 \condvar{Q}{Q_A} = \variance{\delta Q_A}  
    =\variance{Q}-\frac{| \langle Q'\,Q\rangle | ^2}{\variance{Q'}}.  
\end{equation}
By using the commutation relation 
 \begin{equation}
[\delta Q_A, P]=\underbrace{[Q, P]}_{2iN_0} -\alpha\,\underbrace{[Q',P]}_0 ,
\end{equation}
which directly follows from the definition of $\delta Q_A$,
we find that the following uncertainty relation on the beam $(Q,P)$ after
the measurement of $Q'$ holds~:
\begin{equation}
\condvar{Q}{Q_A}\times \variance{P}\ge N_0^2 \label{eq:condvarvar}.
\end{equation}
Using the expression \eqref{eq:ValCondVar}, we obtain
\begin{equation}
 \lvert\langle Q'\,Q\rangle\rvert^2
  \le
    \variance{Q'}\variance{Q} -N_0^2\frac{\variance{Q'}}{\variance{P}}.
\end{equation}
By definition, the EPR beams are maximally correlated and saturate this limit,
which gives
\begin{align}
\langle
Q'\,Q\rangle&=\sqrt{V^2-1}\;N_0 
&
\condvar{Q}{Q_A}&=\frac{N_0}{V}
\end{align}
Since by measuring $Q'$ Alice deduces $Q_A$, and since $Q=Q_A+\delta Q_A$,
the beam $(Q,P)$ is projected onto a $Q$-squeezed state of squeezing
parameter $s=\condvar{Q}{Q_A}/ N_0 = 1/ V $ centered on $(Q_A,0)$.

Alternatively, Alice could measure the quadrature $P'$, yielding the
estimator $P_A=-\alpha P'$, which gives 
\begin{align}
\langle
P'\,P\rangle&=-\sqrt{V^2-1}\;N_0
&
\condvar{P}{P_A}&=\frac{N_0}{V}
\end{align}
Of course, by measuring $P'$,
Alice learns $P_A$ and projects the other beam onto a $P$-squeezed state
centered on $(0,P_A)$ with the same squeezing parameter $s=1/V$.

\subsection{Simultaneous measurement of $Q'$ and $P'$}
\label{sec:SimMeas}

Another possibility for Alice is to measure simultaneously $Q'$ and $P'$.
In this case, her measurement outcomes are more noisy,
so she projects the beam $(Q, P)$ onto a lesser squeezed
state. A crucial point for our protocol is that she prepares a
coherent state if her measurement is balanced in $Q$ and $P$, as we will show below.

\begin{figure} [tbp]
\centerline{\epsfig{file=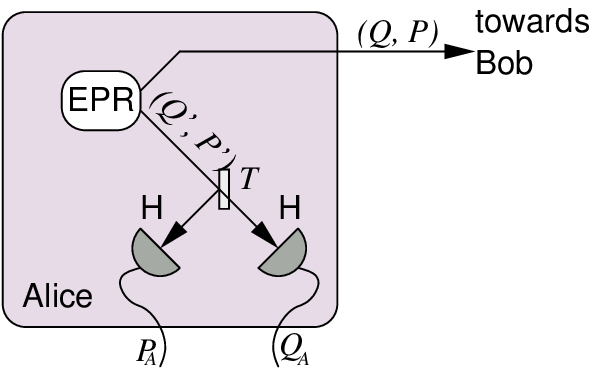, width=8.2cm}} 
\vspace*{13pt}
\fcaption{\label{EPR_RR2}\textbf{Measurement of both
    quadratures.} \textsf{Alice} can measure both quadratures of her beam, as
  explained in the text,  using a beamsplitter of transmission $T$ and two
  homodyne detectors \textsf{H}. She then simultaneously obtains $Q_A$ and $P_A$, which are estimates of $Q$ and $P$.} 
\end{figure}

Denoting as $Q'_A$ and $P'_A$ the values of $Q'$ and $P'$ measurements,
the associated added noises $\delta Q'_A$ and $\delta P'_A$ are defined as
\begin{align}
\delta Q'_A&=Q'-Q'_A   & \delta P'_A&=P'-P'_A,
\end{align}
A possible way to perform such a joint measurement is to split Alice's beam
with a beamsplitter of transmission $T$ (in intensity), measuring separately
each quadrature at each output port of the beamsplitter (see
Fig.~\ref{EPR_RR2}).  Then, $Q'_A$ and $P'_A$ are the best estimators of $Q'$
and $P'$, proportional to the outputs of homodyne detectors placed on each of
the output port.

Since $Q'_A$ and $P'_A$ are known simultaneously, they commute, 
$[Q'_A,P'_A]=0$. Therefore, $[\delta Q'_A,\delta P'_A]=-[Q',P']$ and the noise
variances obey the following inequality:
\begin{equation}
\variance{\delta Q'_A}\variance{\delta P'_A}\ge {N_0}^2. \label{eq:HsbgNoise}
\end{equation}
If this inequality is saturated, that is if Alice makes an optimal joint
measurement, this measure is characterized by the positive number $\mu$,
defined by
\begin{align}
\variance{\delta Q'_A}&=\mu\,N_0&
&\text{and}&
\variance{\delta P'_A}&=\frac1\mu\,N_0.
\end{align}
If the measurement is made with the beamsplitter setup described above, we
have
\begin{align}
\mu&=\frac{1-T}T & \text{or}& 
&T=\frac1{1+\mu}
\end{align}

If $\mu=1$, Alice measures $Q'$ and $P'$ with the same (shot-noise limited)
precision. This case corresponds to a 50:50 beamsplitter ($T=\frac12$). If
$\mu<1$, Alice measures $Q'$ with a sub-shotnoise accuracy. At the limit
$\mu\to0$, Alice measures perfectly $Q'$ but not at all $P'$, since the noise
$\delta P'_A$ needs to be infinite in order to fulfill the Heisenberg inequality
\eqref{eq:HsbgNoise}. This limit corresponds to the perfectly transmitting
beamsplitter ($T=1$), where nothing is reflected to the ``$P$-measuring
port''. If $\mu>1$, the situation is reversed, and Alice measures $P'$ more
accurately than $Q'$. At the limit $\mu\to\infty$, she only measures $P'$,
gaining no information on $Q'$.

Now, from the measured quadratures $Q'$ and $P'$, Alice can again estimate
the correlated quadratures $Q$ and $P$. 
Her best estimate of the state of the beam $(Q,P)$ is given by $(Q_A,P_A)$,
which are now defined simultaneously:
\begin{align}
  Q_A&=\frac{\langle Q\,Q'_A\rangle}{\variance{Q'_A}}Q'_A
      =\frac{\sqrt{V^2-1}}{V+\mu}(Q'-\delta Q'_A)&
 &\text{and}&
  P_A&
      =-\frac{\sqrt{V^2-1}}{V+\frac1{\mu}}(P'-\delta P'_A).
\end{align}
Using
\begin{align}
  Q&=Q_A+\delta Q_A&
  &\text{and}&
  P=P_A+\delta P_A,
\end{align}
with $\delta Q_A$ and $\delta P_A$ defining the noise of the estimators,
the conditional variances can be expressed as
\begin{subequations}\label{eq:squeezval}
\begin{align}
  \condvar{Q}{Q_A} = \variance{\delta Q_A}&=\variance{Q}-\frac{\langle
    Q\,Q_A\rangle^2}{\variance{Q_A}}
  =\left(V-\frac{(V^2-1)}{(V+\mu)}\right)N_0\nonumber\\
  &=\frac{\mu V + 1}{V+\mu}N_0
\end{align}
and
\begin{align}
 \condvar{P}{P_A} = \variance{\delta P_A} &=\variance{P}-\frac{\langle
P\,P_A\rangle^2}{\variance{P_A}}
 =\left(V-\frac{(V^2-1)}{(V+\frac1\mu)}\right)N_0\nonumber\\
 &=\frac{V+\mu}{\mu V + 1}N_0=\frac{{N_0}^2}{\condvar{Q}{Q_A}}
\end{align}
\end{subequations}
Said otherwise, the measurement of $Q'$ and $P'$ projects the beam $(Q, P)$ onto
a squeezed state of variances $\condvar{Q}{Q_A}$ and $\condvar{P}{P_A}$. 
Then, it is clear that 
if the measurement is symmetric in $Q'$ an $P'$ (\ie\ if $\mu=1$),
one has $\condvar{Q}{Q_A} = \condvar{P}{P_A} = N_0$ and
the beam $(Q,P)$ is projected onto a coherent state.  
The mean values of the quadratures of the beam $(Q,P)$ are given by $Q_A$ and $P_A$, so things happen 
as if Alice had prepared a randomly displaced squeezed (or coherent) state.

\subsection{Virtual entanglement}

\begin{figure} [tbp]
\centerline{\epsfig{file=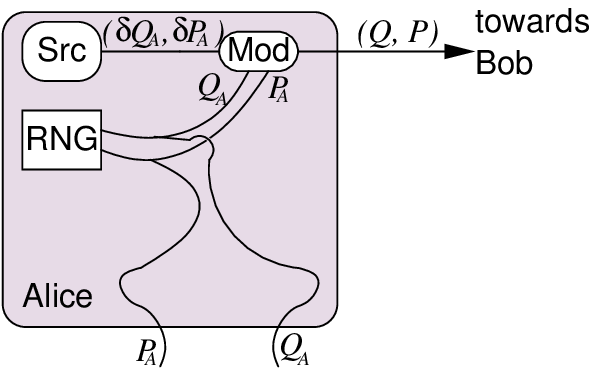, width=8.2cm}} 
\vspace*{13pt} \fcaption{\label{EPR_RR3}\textbf{Equivalent black box.} The
  system sketched in Fig.~\ref{EPR_RR2} is equivalent to this black box. A
  random number generator (\textsf{RNG}) gives two values $Q_A$ and $P_A$.
  A squeezed (or coherent if $s=1$) state source (\textsf{Src}) generates the
  beam $(\delta Q_A,\delta P_A)$, wich is then displaced in phase space by
  $(Q_A,P_A)$ using a modulator (\textsf{Mod}).}
\end{figure}

Let us suppose the EPR source and the measuring apparatus of Alice are hidden
in a black box. The only things coming out of this black box are the values of
$Q_A$ and $P_A$, and the beam $(Q,P)$. This black box is indistinguishable
from an equivalent black box, sketched in Fig.~\ref{EPR_RR3}, where $Q_A$ and
$P_A$ are chosen by the adequate random generator and the beam $(Q,P)$ is in
the displaced squeezed state centered around $(Q_A, P_A)$. Its squeezing
factor is
\begin{equation}
s=\frac{\condvar{Q}{Q_A}}{N_0}=\frac{\mu V +1}{V+\mu},
\end{equation}
and the equations \eqref{eq:squeezval} can be rewritten
\begin{align}
  \condvar{Q}{Q_A}=\variance{\delta Q_A}=&s\,N_0&
  &\text{and}&
  \condvar{P}{P_A}=\variance{\delta P_A}=&\frac{N_0}{s}
\end{align}

The black box with $\mu=0$ and in the case $Q$ and $P$ are randomly interchanged
allows therefore to prepare the randomly displaced squeezed states
that are used in the QKD protocol described in \cite{CLVA00, CIVA01}. 
If we fix $\mu$ to any given value, we realize all of the protocols 
presented in \cite{prl}. In particular,
since $\mu=1$ corresponds to the preparation of a coherent state
($s=1$), the modulated coherent states QKD protocols used in
\cite{prl, GVAWBCG03, GG02} are equivalent to entanglement-based protocols
even if they neither use squeezing nor entanglement. This possibility
to prepare randomly displaced coherent states with an entanglement-based setup
was implicitly present in our previous security studies of individual gaussian attacks on 
reverse reconciliation protocols \cite{GVAWBCG03, GG02}. It is also
useful to extend the Gottesman-Preskill proof of unconditional security
of squeezed-state protocols \cite{GP01} in an attempt to demonstrate the security of 
coherent-state protocols with respect to general attacks \cite{VAIC04}.

We call this possibility \emph{virtual entanglement}: even if Alice does not actually use 
entanglement to create her coherent (or squeezed) states, there exists an equivalent setup (the black box described above) which uses entanglement to
create them.  This relies on the fact that the outputs of any physical apparatus, including Eve's eavesdropping system, can only depend on the
density matrix of its input (in this case, the beam sent by Alice), and not on the way it was prepared. Cryptographic security is then related not to
the transmission of ``real'' entanglement, but rather to the ability of the quantum channel to transmit entanglement, as we will show below.

\section{Bounding Eve's attack on reverse reconciliation}
\label{ReverseCloning}

\subsection{Entangling cloner}

\begin{figure} [htbp]
\centerline{\epsfig{file=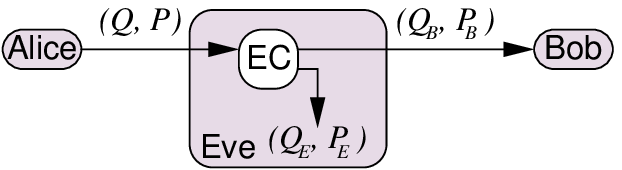, width=8.2cm}} 
\vspace*{13pt}
\fcaption{\label{EntClon} \textbf{Eve's attack on reverse reconciliation.} To
  attack a reverse reconciliation QKD protocol, \textsf{Eve} uses an
  entangling cloner (\textsf{EC}). It takes \textsf{Alice}'s beam $(Q,P)$ as
  input and produces two entangled outputs, $(Q_E,P_E)$, which is kept by
  \textsf{Eve}, and $(Q_B,P_B)$, which is sent to \textsf{Bob} through a perfect line.} 
\end{figure}

To eavesdrop a reverse reconciliation scheme, Eve needs to guess the results
of Bob's measurement. We will call {\it entangling cloner} a system allowing
her to do so, because this kind of system can be described as a cloner creating
two entangled outputs, Eve keeping one of them and sending the other one to
Bob (see Fig.~\ref{EntClon}). Here $(Q,P)$ are the input quadratures of the entangling cloner and
$(Q_B,P_B)$, $(Q_E,P_E)$ the quadratures of its two outputs. A good entangling
cloner should minimize the conditional variances \cite{qnd,qnd2} $V_{Q_B|Q_E}$ and
$V_{P_B|P_E}$.

Alice and Bob should assume Eve uses the best possible entangling
cloner, knowing the Alice-Bob channel quality. This channel can be
described by
\begin{align}
  Q_B&=\sqrt{G_Q}\,(Q+\delta Q_B) &
&\text{and}&
  P_B&=\sqrt{G_P}\,(P+\delta P_B),
\end{align}
with
\begin{align}
  \left<\delta Q_B^2\right>&=\chi_Q \; N_0, 
& \left<\delta P_B^2\right>&=\chi_{P} \; N_0 &
&\text{and}&
  \left<Q\,\delta Q_B\right>=\left<P\,\delta P_B\right>&=0
\end{align}
\subsection{Heisenberg inequalities on Alice's and Eve's conditional variances}
\label{SecHsbgVarCond} For reverse reconciliation protocols, Alice
needs to evaluate $Q_B$. Her estimator can be noted $\beta Q_A$, with
$\beta=\frac{\langle Q_A\,Q_B\rangle}{\variance{Q_B}}
=\frac{V-s}{\sqrt{G_Q}(V+\chi_Q)}$.  Eve's estimator for $P_B$ will be
$P_E$. The error of these estimators are
\begin{align}
  Q_{B|A}     &=Q_B-\beta  Q_A&
 &\text{and}&
  P_{B|E}&=P_B-P_E.
\end{align}
The commutator of these two quantities is then equal to 
\begin{equation} 
[Q_{B|A},P_{B|E}]
= [Q_B,P_B] -\beta\underbrace{[Q_A,P_B]}_0
            -\underbrace{[Q_B,P_E]}_0
            +\beta\underbrace{[Q_A,P_E]}_0.  
\end{equation} 
We have therefore $[Q_{B|A},P_{B|E}]=[Q_B,P_B]=2iN_0$.
This commutation relation leads to the following inequality on conditional
variances:
\begin{align}
\label{HsbgEA}
 V_{Q_B|Q_A}V_{P_B|P_E}&\ge N_0^2&
         &\text{and}&
  V_{P_B|P_A}V_{Q_B|Q_E} &\ge      N_0^2,
\end{align}
the second inequality being obtained by exchanging the roles of $Q$ and $P$.
%
These inequalities mean that Alice and Eve cannot jointly
know more about Bob's field than allowed by the Heisenberg principle. 

\subsection{Alice's conditional variance}

Alice's conditional variance on $Q_B$ is
\begin{align}
\condvar{Q_B}{Q_A}
       &=\left<Q_B^2\right>-\frac{\left<Q_A\,Q_B\right>^2}{\left<Q_A^2\right>}
           =G_QVN_0+G_Q\chi_QN_0 -G_QVN_0+G_QsN_0\nonumber\\
           &= G_Q(\chi_Q + s)N_0
            \label{VqBA}
\end{align}
A similar calculation leads to the symmetric relation
 \begin{equation}
 \label{VpBA} V_{P_B|P_A}=G_P (\chi_P + \tfrac1s)N_0.
 \end{equation}
These conditional variances depend on the amount of squeezing $s$ Alice 
generates with her black-box. Therefore, the constraint on squeezing
$\frac1V<s<V$ gives us the minimal values of these conditional variances
\bea
  V_{P_B|P_A}&\ge& V_{P_B|P_A,\text{min}}=G_P (\chi_P+\tfrac{1}V)N_0   \\
  V_{Q_B|Q_A}&\ge& V_{Q_B|Q_A,\text{min}}=G_Q ( \chi_Q+\tfrac{1}V)N_0
\label{VBAmin}
\eea

\subsection{Eve's conditional variance}
\label{VarianceCondEve}

The output-output correlations of an entangling cloner, described e.g. by
$V_{P_B|P_E}$, should only depend on the density matrix of the field $(Q,P)$
at its input, and not on the way this field was built. The inequality
(\ref{HsbgEA}) has thus to be fulfilled for every physically allowed value of
$V_{Q_B|Q_A}$, given the density matrix of the field $(Q,P)$. 
Since this field is gaussian, its density matrix is uniquely defined by its
covariance matrix, \ie by the parameters
$\variance{Q}=\variance{P}=V\,N_0$ and $\langle Q\;P\rangle=0$, and we have to
consider all possible black-boxes (those of Fig.~\ref{EPR_RR2} as well as those of Fig.~\ref{EPR_RR3}). 
In order to bound Eve's knowledge by using Eq.(\ref{HsbgEA}), we thus have to
use the tightest limit on $V_{Q_B|Q_A}$, which is given by
$V_{Q_B|Q_A,\text{min}}$ according to (\ref{VBAmin}). Obviously the same
reasoning holds for $V_{P_B|P_A}$, with the corresponding tightest limit
$V_{P_B|P_A,\text{min}}$.

We have then
 \beq
 V_{Q_B|Q_E} \ge        V_{Q_B|Q_E,\text{min}}=\frac{N_0}{G_P(\chi_P+1/V)}
 \label{VBExmin}
\end{equation}
 and, similarly
 \beq
  \label{VBEpmin}
 V_{P_B|P_E}\ge      V_{P_B|P_E,\text{min}}=\frac{N_0}{G_Q(\chi_Q+1/V)}
\end{equation}
If one of these inequalities was violated and if Alice had prepared her field with an 
EPR-beams based black-box, then Eve and Bob would be able to
make a joint measurement of the field $(Q',P')$ with a better accuracy than allowed by the Heisenberg uncertainty limit. 

\subsection{Implementation of the entangling cloner}
\label{VraiCloneurInverse}

\begin{figure} [htbp]
\centerline{\epsfig{file=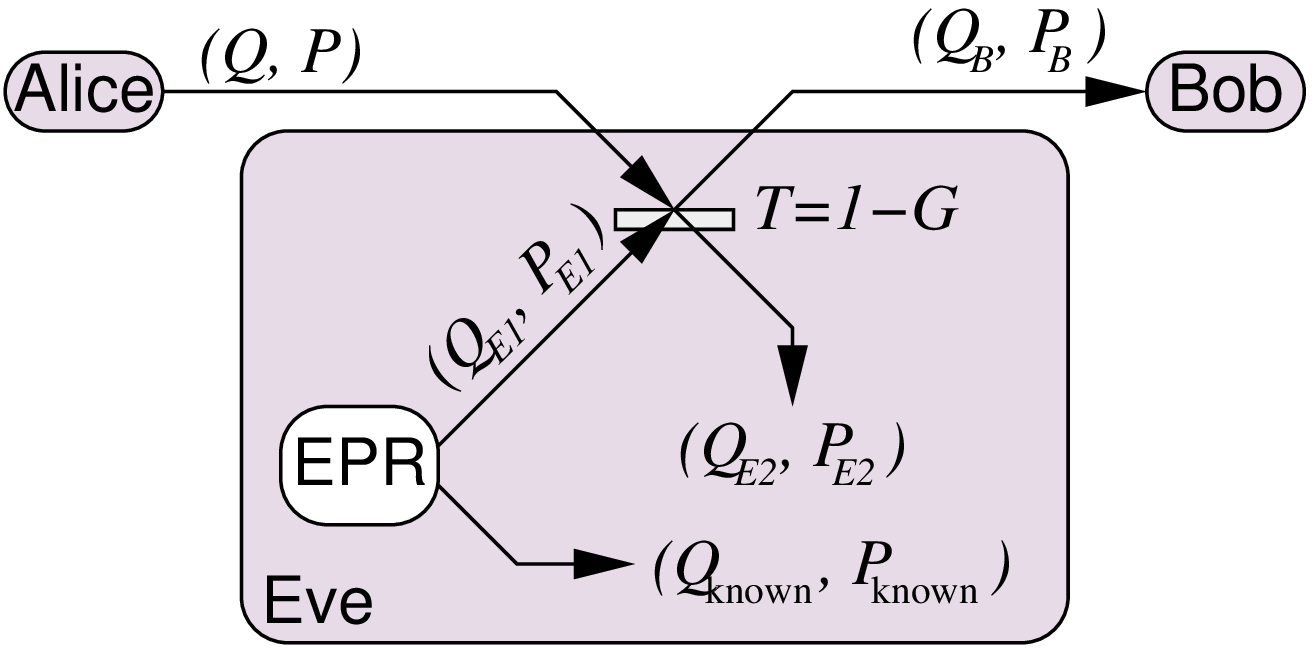, width=8.2cm}} 
\vspace*{13pt} \fcaption{\label{EntClon2} \textbf{Implementation of an
    entangling cloner} for $G<1$. \textsf{Eve} uses a beamsplitter of transmission $T=1-G$ to inject into the line a
    partially known noise $(Q_{E1},P_{E1})$ generated with an EPR source
  (\textsf{EPR}). She keeps
    the other output $(Q_{E2}, P_{E2})$ of the beamsplitter which, combined
    with her knowledge $(Q_{\text{known}}, P_{\text{known}})$ on the injected
    noise, gives her an estimate of \textsf{Bob}'s beam $(Q_B,P_B)$.}
\end{figure}

In a practical QKD scheme, Alice and Bob will give the same roles to $Q$ and
$P$. Assuming therefore that $G_Q=G_P=G$ and $\chi_Q=\chi_P=\chi$, the two
bounds above reduce to a single one, and it is possible to explicitly describe
an entangling cloner achieving this limit. We will consider here only the case
where $G<1$, but the limit is tight for any $G$. The entangling cloner can
then be sketched as shown in Fig~\ref{EntClon2}: Eve uses a beamsplitter
with a transmission $G$ to split up part of the Alice-Bob transmitted signal,
and she injects into the other input port a field $E1$, with the right
variance to induce a noise of variance $G\chi N_0$ at Bob's end. One has
therefore:
\begin{align}
  \left<Q_{E1}^2\right> &= \frac{G \chi N_0}{1-G} &
  \left<P_{E1}^2\right> &= \frac{G \chi N_0}{1-G}
\end{align}
 Eve should know the maximum about this injected field $E1$, and
 will therefore use an half-pair of EPR-correlated beams, so that
she does perform an ``entangling" attack. We can then write
 \beq
  Q_{E1}=Q_{\text{known}}+Q_{\text{unknown}}
 \end{equation}
 where $Q_{\text{known}}$ stand for Eve's best estimation of
 $Q_{E1}$, given by the measure of its brother-beam, and
 $Q_{\text{unknown}}$ stand for the noise she cannot know.
We have
 \bea
  \left<Q_{\text{unknown}}^2\right>&=&\frac{N_0^2}{\left<Q_{E1}^2\right>}=\frac{(1-G)N_0}{G \chi}\\
  \left<Q_{\text{known}}^2\right>&=&\left<Q_{E1}^2\right>- \left<Q_{\text{unknown}}^2\right>
 \eea
Eve also use an output port of the beamsplitter to measure the
field $E2$, which gives her information about the input field:
 \beq
 Q_{E2}=\sqrt G\,Q_{E1}-\sqrt{1-G}\,Q.
 \end{equation}
 She can cancel a part of the noise induced by $E1$ by subtracting
 the part proportional to $Q_{\text{known}}$. Thus she knows
  \beq
  Q'_{E2}=\sqrt G\,Q_{\text{unknown}}-\sqrt{1-G}\,Q.
  \end{equation}
 We also have
 \beq
  Q_B=\sqrt G\,Q+\sqrt{1-G}\,Q_{E1}.
 \end{equation}
 where Eve already knows the part proportional to $Q_{\text{known}}$,
 injected with $Q_{E1}$ and she only needs to guess
  \beq
  Q'_B=\sqrt G\,Q+\sqrt{1-G}Q_{\text{unknown}}
 \end{equation}
 from $Q'_{E2}$.
 We have therefore
 \beq
  V_{Q_B|Q_{E1},Q_{E2}}=V_{Q'_B|Q'_{E2}}.
 \end{equation}
The calculation of the quantities $\left<Q_B^{\prime2}   \right>$,
$\left<Q_{E2}^{\prime2}\right>$, $\left<Q'_{E2}Q'_B     \right>$
leads straightforwardly to the conditional variance
\beq
 V_{Q'_B|Q'_{E2}}  = \frac{N_0}{G\chi+G/V} = V_{Q_B|Q_E,\text{min}}
\end{equation}
showing that the entangling cloner does reach the lower limit 
of Eqs. (\ref{VBExmin}) and (\ref{VBEpmin}).

\section{Security of reverse-reconciliation based quantum cryptography}
\label{security}

  \subsection{Tolerable noise}
 In a reverse reconciliation protocol, Eve's power is
 limited by the values of $V_{Q_B|Q_E,\text{min}}$ and
 $V_{P_B|P_E,\text{min}}$ given by Eqs. (\ref{VBExmin}) and (\ref{VBEpmin}). In a security analysis, 
we have to assume that a ``perfect'' Eve is able to reach this limit, that is,
\begin{subequations}
\label{VBE}
\bea
   V_{Q_B|Q_E}&=&V_{Q_B|Q_E,\text{min}}=\frac{N_0}{G_P(\chi_P+1/V)}\\
   V_{P_B|P_E}&=&V_{P_B|P_E,\text{min}}=\frac{N_0}{G_Q(\chi_Q+1/V)}
\eea
\end{subequations}
On Alice's side, the relevant conditional variances are given by
Eqs. (\ref{VqBA}) and (\ref{VpBA}). Alice's and Eve's conditional variances
can be converted into mutual informations by using Shannon's formula\cite{Sha48}. For
the quadrature $Q$, we have
 \begin{align}
  I_{BA}^Q&=\frac12\log_2\frac{\left<Q_B^2\right>}{V_{Q_B|Q_A}}  &
  I_{BE}^Q&=\frac12\log_2\frac{\left<Q_B^2\right>}{V_{Q_B|Q_E}}
 \end{align}
while, for the quadrature $P$, we have
 \begin{align}
  I_{BA}^P&=\frac12\log_2\frac{\left<P_B^2\right>}{V_{P_B|P_A}}  &
  I_{BE}^P&=\frac12\log_2\frac{\left<P_B^2\right>}{V_{P_B|P_E}}
 \end{align}
Following \cite{CK78, Maurer}, we know that a sufficient condition
for reverse reconciliation to give a non-zero secret key rate
is $I_{BA}^Q > I_{BE}^Q$ (for the $Q$ quadrature) or $I_{BA}^P > I_{BE}^P$
(for the $P$ quadrature). In terms of conditional variances, this translates
into
\begin{align}
\label{ComparVcnd}
 V_{Q_B|Q_E} &> V_{Q_B|Q_A}&
&\text{ or } &
V_{P_B|P_E} >& V_{P_B|P_A}
\end{align}
Using Eqs. (\ref{VqBA}), (\ref{VpBA}) and (\ref{VBE}), 
we obtain  (sufficient) conditions for the security
of a reverse-reconciliation based protocol
 \begin{align}
  (G_Q\chi_Q+G_Qs)(G_P\chi_P+\frac{G_P}V) &<  1 &
  &\text{or}&
  (G_P\chi_P+G_Ps)(G_Q\chi_Q+\frac{G_Q}V) &<  1.
\end{align}
For simplicity reasons, we will assume in the following that all equations
are symmetric in $Q$ and $P$, in particular $G_Q=G_P=G$ and
$\chi_Q=\chi_P=\chi$,\footnote{Any experimental implementation of this
protocol should however estimate these parameters from
statistical tests, which are likely not to be exactly symmetric.} \ 
so that these conditions simplify into:
\beq
  (G\chi+G s)(G \chi + G/V) < 1.
\label{CondCryptoGen}
\end{equation} 
 This condition can be rewritten by using the definition
$\chi = \chi_0 + \varepsilon$, where $\chi_0=\frac{1-G}G$ is the loss-induced ``vacuum noise'' and $\varepsilon$ is the excess
noise\footnote{Strictly speaking, $\varepsilon$ corresponds to the excess noise
only in the usual case of losses, where $G\le1$.}\ , giving
\begin{equation}
 \label{eq:CondRReps}
 [1-G(1-s-\varepsilon)][1-G(1-\tfrac1V-\varepsilon)] < 1.
\end{equation}
Since $s\le 1$ and $V>1$, this condition is always fulfilled for
$\varepsilon=0$, \ie when the noise only originates from losses. This holds for arbitrary high losses ($G\to 0$) 
and even for coherent state protocols ($s=1$). Therefore, reverse reconciliation provides a simple way 
to extend the coherent state protocol of ref.
\cite{prl} into the high-loss regime.

\begin{figure}[tbp]
\centerline{\epsfig{file=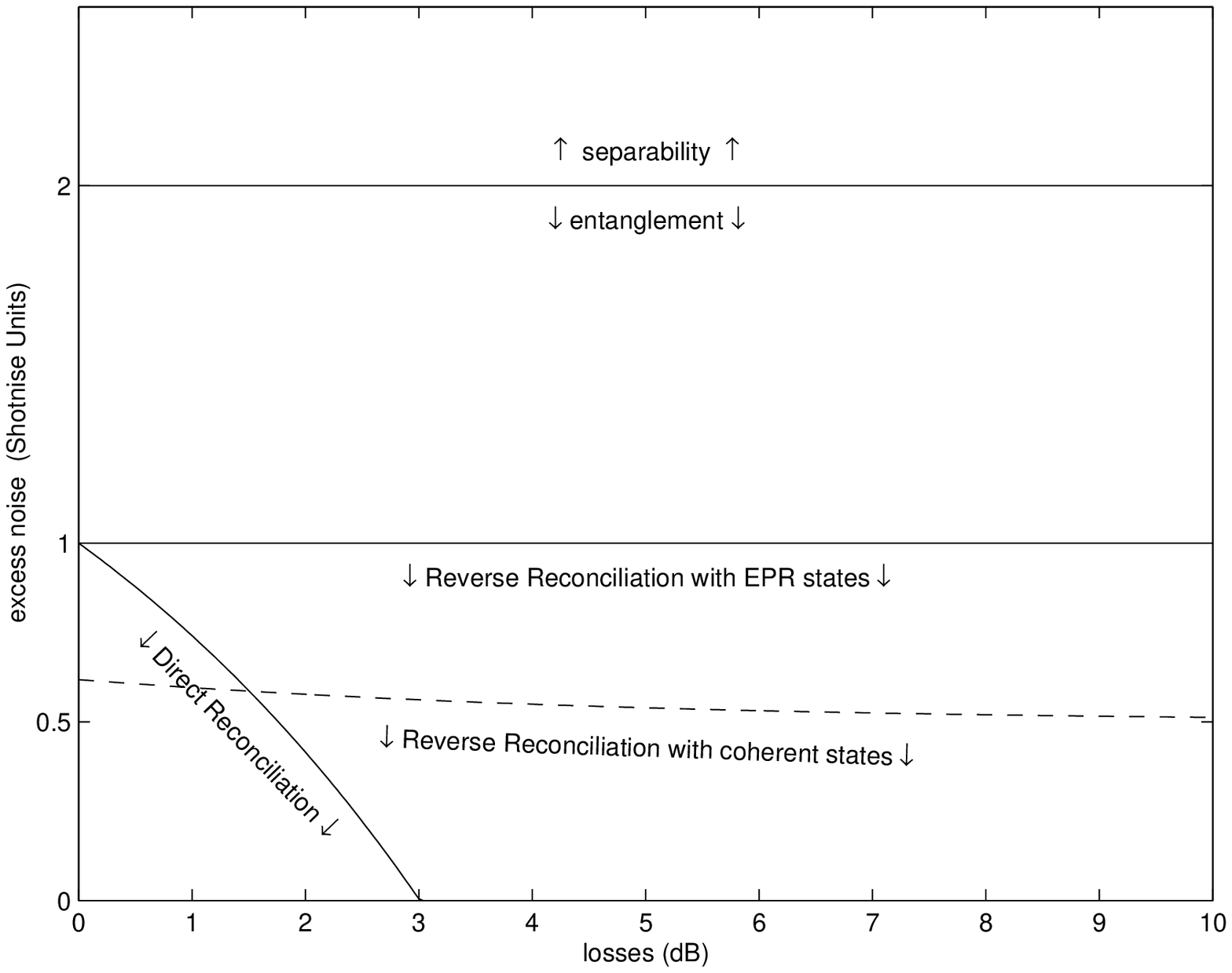, width=8.2cm}} 
\vspace*{13pt} \fcaption{\label{Graphe}\textbf{Tolerable excess noise
    $\varepsilon$ as a function of the losses} at the high modulation limit
  ($V\gg 1$).  The RR limit is given by Eq.~\eqref{eq:GenEpsilonmax}.
  It reduces to $\varepsilon_{\max}^{\text{EPR}}=1$ for EPR states (or maximal
  squeezing) at the high modulation limit ($s=\tfrac1V\to 0$), and to
  Eq.~\eqref{eq:CohEpsilonmax} for coherent states (dashed line). 
  The DR security limit defined in Eq.~\eqref{eq:dirlim} implies that DR 
  is more robust against excess noise than coherent state RR in the low losses regime.  
The entanglement limit given by Eq.~\eqref{eq:entlim}, \ie $\varepsilon=2$, is well above the previous security limits.  
In the region $1<\varepsilon<2$,
no QCV cryptographic protocol is known, although entanglement is present.}
\end{figure}

Finally, one can show that squeezed state protocols are more robust against
excess noise than coherent state protocols. Indeed, by solving Eq.~(\ref{eq:CondRReps}), we get
\begin{align}
  \label{eq:GenEpsilonmax}
  \varepsilon&<\varepsilon_{\max}&
  &\text{with}&
  \varepsilon_{\max}=1-\tfrac1V \underbrace{ -\tfrac1G-\tfrac12(s-\tfrac1V) +
     \sqrt{\tfrac1{G^2}+\tfrac14(s-\tfrac1{V})^2} }_{\le 0} <1
\end{align}
It is easy to check that this upper limit on $\varepsilon$
is less stringent for low values of $s$, \ie for strong squeezing.
When the squeezing is
maximum ($s=\tfrac1V$), we get $\varepsilon_{\max}=1-\tfrac1V$.
Note also that, in the limit of high losses ($G\to 0$), we have
$\varepsilon_{\max}=1-\tfrac12 (s+\tfrac1V)$. The maximum tolerable excess noise  is shown in 
Fig.~\ref{Graphe} as a function of the losses in the limiting case of high modulation ($V\to\infty$).

\subsection{Secret information rates (EPR vs coherent beams)}
\label{SecIRates}
The condition (\ref{CondCryptoGen}) can directly be translated into 
a secret information rate by using Shannon's formula (in the case where
everything is symmetric in $Q$ and $P$) \cite{Sha48}
 \begin{align}
  I_{BA}&=\tfrac12\log_2\frac{\left<Q_B^2\right>}{V_{B|A}}  &
  I_{BE}&=\tfrac12\log_2\frac{\left<Q_B^2\right>}{V_{B|E}}  \\
        &=\tfrac12\log_2\frac{V+\chi}{s+\chi} &
        &=\tfrac12\log_2[(GV+G\chi)(G\chi+G\tfrac1V)]
 \end{align}
The RR secret information rate is therefore
 \begin{equation}
  \Delta I=I_{BA}-I_{BE}=\tfrac12\log_2\frac{V_{B|E}}{V_{B|A}}
  =\tfrac12\log_2\frac{1}{\left(G\chi+G\frac{1}{V}\right)(G\chi+Gs)}
  \label{DeltaI}
 \end{equation}
and it is strictly positive if the security condition (\ref{CondCryptoGen}) is fulfilled.

%

Let us compare the cases where Alice uses EPR or coherent beams.
If Alice measures only one quadrature of an EPR beam (or modulates a
maximally squeezed beam compatible with the total variance $V$), we have
$s=1/V$ and $\varepsilon_{\max}^{\text{EPR}}=1-\tfrac1V$. 
Alice and Bob gain shared information only every second transmission since they
don't always choose the same measurement basis\footnote{We suppose that Alice and Bob do not have a quantum memory available.}. Therefore,
 \begin{align}
  \Delta I_{\text{EPR}}
   =\frac14\log_2\frac{1}{\left(G\chi+G\frac{1}{V}\right)^2} 
   &=\frac12\log_2\frac{1  }{      G\chi+G\frac{1}{V}         }\\
  \label{DeltaIEPR}
   &=\frac12\log_2\frac{1  }{1-G\left(1-\tfrac{1}{V}-\varepsilon\right)}
\end{align}
In contrast, for coherent beams, we have $s=1$ and 
\begin{equation}
\label{eq:CohEpsilonmax}
\varepsilon_{\max}^{\text{coh}}=\tfrac12-\tfrac1{2V}-\tfrac1G+
\sqrt{\tfrac1{G^2}+\tfrac14(1-\tfrac1V)^2}. 
\end{equation}
The mutual informations are not dependent of the basis choice (we do not get this prefactor 1/2), so we have
\bea
  \Delta I_{\text{coh}}&=&\frac12\log_2\frac1{\left(G\chi+G\frac1{V}\right)(G\chi +G)}\\
  \label{DeltaIcoh}
  &=&\Delta I_{\text{EPR}}-\frac12\log_2(1+G\varepsilon)
\eea
Since the excess noise $\varepsilon$ is positive, we obtain
  \beq
  \Delta I_{\text{coh}}\le\Delta I_{\text{EPR}}.
  \end{equation}
 Both secret rates become equal if and only if the noise only comes from
 losses ($\varepsilon=0$ and $G\le 1$). As in \cite{prl}, the use of entanglement or squeezing does not improve the secret
 rate for losses only, and it becomes advantageous only in the presence of excess noise.
\subsection{Strong losses limit}
\label{FortesPertes}

Assuming strong losses ($G\ll1$), Eqs. \eqref{DeltaIEPR} and \eqref{DeltaIcoh} tend to
\begin{align}
\Delta I_{\text{EPR}}&\simeq\frac{G}{2\ln2}\left(1-\tfrac1V-\varepsilon\right)&
\Delta I_{\text{coh}}&\simeq\frac{G}{2\ln2}\left(1-\tfrac1V-2\varepsilon\right)
\end{align} 
In the case where there is no excess noise ($\varepsilon=0$), both rates are 
equal, as we just said, and we get $ \Delta I_{\text{EPR,losses}}
= \Delta I_{\text{coh,losses}}$.
If there is some excess noise in the line, one sees that the
reverse reconciliation protocol is secure as long as as $\varepsilon < \tfrac12(1-\tfrac1V) \sim 1/2$ 
for coherent states, and $\varepsilon < 1-\tfrac1V \sim 1$ for EPR beams. This shows again that it is always possible to use coherent states regardless
the line losses, though EPR beams make the scheme more robust against excess noise.

Now, we may compare the secret key rate of the RR coherent-state protocol with
BB84's net key rate in the case of a lossy errorless channel, which is
$\frac12G\bar n$ with $\bar n=1$ for single photons and $\bar n\ll1$ for weak
coherent pulses. Taking for instance a 100~km line with 20~dB loss ($G=0.01$)
and a reasonable modulation ($V\simeq10$), the secret key rate is $\Delta
I=6.5\cdot10^{-3}$~bit/symbol for a RR coherent-state protocol. For the same
parameters, the secret key rate for BB84 with an ideal single-photon source
would be at best $5\cdot10^{-3}$~bit/time slot, and one order of magnitude
smaller using attenuated light pulses with $\bar n=0.1$, even with perfect
detectors (this corresponds to a very recent experimental realization of BB84 \cite{kosaka}). 
Thus, our reversed-reconciliation QCV protocol has, in principle, a comparable efficiency to that of ideal BB84 (for strong losses and no excess noise). 
In particular, with a ``symbol rate'' of a few MHz, which should be easy to achieve, the theoretical QCV secret key rate after 100 km would be more than
10 kbits/sec.

We must stress, however, that in order to achieve this
rate, better reconciliation protocols than those available today should be
developed. In their current state, the reconciliation procedures cannot
extract a single secret bit in such a high-loss regime (the highest loss that
can be tolerated in the first experimental demonstration of QCV quantum
cryptography is about 3.1~dB \cite{GVAWBCG03}). Indeed, for the values of the
parameters above, the information between Alice and Bob is
$I_{AB}=6.2\cdot10^{-2}$ bit/symbol, which is one order of magnitude larger
than $\Delta I$. Hence, the required reconciliation efficiency should be
larger than 90 percent in a regime where the information content ($I_{AB}$) is
of a few hundredth of bit per symbol (or, in other words, when the
signal-to-noise ratio does not exceed about $-10$ dB).

\section{Entanglement versus security criteria}
\label{entanglement-security}

\subsection{Virtual entanglement criterion}
If the channel between Alice and Bob is too noisy, the \emph{virtual
  entanglement} between $(Q_B,P_B)$ and $(Q',P')$ will be destroyed. The
threshold at which this happens can be calculated using the Duan--Simon entanglement criterion for bivariate gaussian states\cite{DGCZ00a, Simon00}.
This criterion, expressed by the equation (17) of \cite{DGCZ00a}, is
\begin{equation}
  \label{eq:SimpEPRtxt}
  (V-1)(V_B-1)<C^2,
\end{equation}
where
\begin{gather}
V_B\,N_0=\variance{Q_B}=\variance{P_B}=G(V+\chi)\,N_0\\
C\,N_0=\langle Q'\,Q_B\rangle=-\langle P'\,P_B\rangle=\sqrt{G(V^2-1)}\,N_0
\end{gather}
In our case, this leads to
\begin{align}
  G(V-1)(V-1+\varepsilon)&<G(V-1)(V+1)&
   &\Leftrightarrow&
  \varepsilon&<2 \label{eq:entlim}
\end{align}
Therefore, virtual entanglement is present as soon as there is non-zero
modulation ($V>1$) and non-zero transmission ($G>0$), provided that the
excess noise of the channel is smaller than twice the shot-noise limit.

\subsection{Security criteria}


The security limit against gaussian individual attacks of the QKD protocols
discussed in \cite{prl, GVAWBCG03, CLVA00, CIVA01, GG02} are simply obtained
by comparing conditional variances. For direct protocols \cite{prl}, an
argument linked to cloning leads to the limit \cite{prl, CLVA00, CIVA01}
\begin{align}
 \chi&<1&
 &\Leftrightarrow&
 \varepsilon&<2-\frac1G,\label{eq:dirlim}
\end{align}
which ensures that the inequality \eqref{eq:entlim} is fulfilled.
For reverse protocols, the inequality \eqref{eq:GenEpsilonmax}
cannot be fulfilled if $\varepsilon>1$ so that the entanglement
condition $\varepsilon<2$ is also always fulfilled when reverse
reconciliation is possible. This situation is
summarized in Fig.~\ref{Graphe}, where the entanglement limit 
is compared with the DR and RR security limits.
The figure makes clear that 
the DR and RR cryptographic security thresholds lie well within the entanglement region, 
where the channel is able to distribute quantum entanglement. 
This holds even if no entangled beams are physically implemented. 

%

It is worth noting that the entanglement threshold is known
to coincide, physically, with an intercept-and-resend attack \cite{horodecki}. 
In other words, at the point where the joint state of Alice and Bob becomes
separable ($\varepsilon=2$), there exists an explicit intercept-and-resend attack, 
so that obviously no protocol can be secure. The gap between the entanglement 
condition \eqref{eq:entlim} and the security limits
\eqref{eq:dirlim} and \eqref{eq:CondRReps} corresponds to a region where the known DR and
RR protocols are insecure with respect to gaussian attacks, 
though intercept-and-resend attacks cannot be used yet. 
It is presently unknown whether improved protocols may be devised, 
that would remain secure against gaussian attacks in this region.

\section{Conclusion}

In this paper we have shown that reverse reconciliation protocols
can be used to extract a secret key from the exchange of coherent, squeezed
or EPR beams between Alice and Bob. The key is secure against individual gaussian 
attacks regardless the transmission of the optical line between Alice and Bob, 
provided that the excess noise (\ie the noise beyond the
loss-induced vacuum noise) is not too large. Squeezing or entanglement allow 
these protocols to tolerate a larger amount of excess
noise, but they are not absolutely required. We have also shown that the QCV protocols 
based on gaussian displaced squeezed or coherent
states \cite{prl, GVAWBCG03, CLVA00, CIVA01, GG02} are equivalent to 
entangled-beams based protocols, and that the security limits of
these protocols are more severe than the entanglement limit of the 
equivalent entanglement-based protocol. This result is certainly
compatible with---and even supports---the idea that they may be 
unconditionally secure \cite{VAIC04}.

The difference between the entanglement condition and the
security limits in RR or DR shows that our
protocols do not use the full available entanglement.  
In principle, procedures based either on quantum entanglement distillation
\cite{Dal00,GP01} or on classical advantage distillation \cite{gisinwolf} 
can exploit the entanglement up to its ultimate limit. However, it should 
be noticed that such protocols are either much more difficult
to implement (quantum entanglement distillation) than the ones we have 
considered here, or have extremely low practical secret bit
rates (classical advantage distillation). It remains an open question 
to determine whether the gap  between our security threshold and
the entanglement threshold is due to the restricted observables we can 
measure through homodyne detection, 
or to the reconciliation procedure used to extract the bits, or perhaps to another factor.

\nonumsection{Acknowledgments}
\noindent
FG acknowledges support from the Belgian National Fund for Scientific Research.
NJC acknowledges financial support from the Communaut\'e Fran\c caise de
Belgique under grant ARC 00/05-251, 
from the IUAP programme of the Belgian governement under grant V-18, 
and from the EU under project RESQ (IST-2001-35759).
This work has been partly funded by the IST / FET / QIPC project ``QUICOV". 

\bigskip

\nonumsection{References}


\noindent

\begin{thebibliography}{000}

\bibitem{braunstein-pati} S.L. Braunstein and A.K. Pati, {\it Quantum Information Theory with Continuous Variables}, (Kluwer Academic, Dordrecht, 2003).

\bibitem{gisin-rmp} N. Gisin, G. Ribordy, W. Tittel, and H. Zbinden, Rev. Mod. Phys. {\bf 74}, 145 (2002).

\bibitem{prl} F. Grosshans and Ph. Grangier,  Phys. Rev. Lett. {\bf 88} 057902 (2002); see also e-print quant-ph/0109084.

\bibitem{GVAWBCG03} F. Grosshans, G. Van Assche, J. Wenger, R. Brouri,
N.J. Cerf and Ph. Grangier, Nature {\bf421}, 238 (2003).


\bibitem{epr}  A. Einstein, B. Podolsky, and N. Rosen,
Phys. Rev. {\bf 47}, 777 (1935).

\bibitem{CLVA00} N.J. Cerf, M. L\'evy, and G. Van Assche, Phys. Rev. A {\bf 63},
052311 (2001); see also e-print quant-ph/0008058.


\bibitem{CIVA01} N.J. Cerf, S. Iblisdir and G. Van Assche, Eur. Phys. J. D {\bf 18}, 211 (2002); see also e-print quant-ph/0107077.


\bibitem{cerf} 
N.J.~Cerf, A. Ipe, and X. Rottenberg, Phys. Rev. Lett. {\bf 85}, 1754 (2000);
see also e-print quant-ph/9909037.
\bibitem{cerf2}
N.J.~Cerf and S.~Iblisdir, Phys. Rev. A {\bf 62}, 040301(R) (2000);
see also e-print quant-ph/0005044.

\bibitem{GG01} F. Grosshans and Ph. Grangier, Phys. Rev. A {\bf 64} 010301(R) (2001); see also e-print quant-ph/0012121.

\bibitem{GG02} F. Grosshans and Ph. Grangier, Proc. 6$^{\text{th}}$ Int. Conf.
  on Quantum Communications, Measurement, and Computing (QCMC'02), Rinton
  Press, December 2002; see also e-print quant-ph/0204127

\bibitem{Dal00} L.-M. Duan, G. Giedke, J. I. Cirac, and P. Zoller, Phys. Rev. Lett. {\bf 84}, 4002 (2000).

\bibitem{silb} Ch. Silberhorn, T. C. Ralph, N. Lütkenhaus, and G. Leuchs, 
Phys. Rev. Lett. {\bf 89}, 167901 (2002). 
                                       
\bibitem{BB84} C. Bennett and G. Brassard, Proc. of IEEE International
  Conference on Computers, Systems, and Signal Processing, Bangalore, India
  (IEEE, New-York, 1984), p. 175

\bibitem{Eke91} A.K. Ekert, Phys. Rev. Let. {\bf 67}, 661 (1991).
  
\bibitem{BBM92} C.H. Bennett, G. Brassard and N.D. Mermin, Phys. Rev. Let.
  {\bf 68}, 557 (1992).

\bibitem{VACC01} G. Van Assche, J. Cardinal and N.J. Cerf,
e-print cs.CR/0107030.


\bibitem{CK78} I. Csiszar and J. K\"orner, IEEE Trans. Inf. Theory {\bf 24}, 339 (1978).

\bibitem{Maurer} U.~Maurer, IEEE Trans. Inf. Theory {\bf 39}, 733 (1993).
  

  
\bibitem{qnd} J.-Ph. Poizat, J.-F. Roch and Ph. Grangier, Ann. Phys. (Paris), {\bf 19}, 265 (1994).

\bibitem{qnd2} Ph. Grangier, J.-A. Levenson and J.-Ph. Poizat, Nature {\bf 396}, 537 (1998).

  
\bibitem{GP01} D. Gottesman and J. Preskill, Phys. Rev. A {\bf 63},
022309 (2001); see also e-print quant-ph/0008046.

\bibitem{VAIC04} S. Iblisdir, G. Van Assche, and N. J. Cerf, article in
  preparation.


\bibitem{Sha48} C.E. Shannon, Bell Syst. Tech. J. {\bf 27}
623-656(1948).

\bibitem{kosaka}  H. Kosaka, A. Tomita, Y. Nambu, T. Kimura, and K. Nakamura,
e-print quant-ph/0306066.

\bibitem{DGCZ00a} L.-M. Duan, G. Giedke, J. I. Cirac and P. Zoller,
Phys. Rev. Lett. {\bf 84}, 2722 (2000).

\bibitem{Simon00} R. Simon, Phys. Rev. Lett. {\bf 84}, 2726 (2000).

\bibitem{horodecki} M. Horodecki, P. W. Shor, and M. B. Ruskai, to appear
in Rev. Math. Phys.; see also quant-ph/0302031.

\bibitem{gisinwolf} N. Gisin and S. Wolf, Phys. Rev. Lett. {\bf 83}, 4200 (1999); see also quant-ph/9902048.






\end{thebibliography}
\end{document}